\documentclass[twocolumn,nodate,showpacs,preprintnumbers,
amsmath,amssymb,aps,prl]{revtex4}
\usepackage{amsmath,amssymb}

\begin{document}

\title{Hybrid Quantum Gowdy Cosmology: Combining Loop
and Fock Quantizations}
\author{M. Mart\'{i}n-Benito$^1$, L.J. Garay$^{1,2}$,
and G.A. Mena Marug\'{a}n$^1$} \affiliation{$^1$
Instituto de Estructura de la Materia, CSIC, Serrano
121, 28006 Madrid, Spain\\$^2$ Departamento de
F\'{i}sica Te\'{o}rica II, Universidad Complutense de
Madrid, 28040 Madrid, Spain}

\begin{abstract}
We quantize an inhomogeneous cosmological model using
techniques that include polymeric quantization. More
explicitly, we construct well defined operators to
represent the constraints and find the physical
Hilbert space formed by their solutions, which
reproduces the conventional Fock quantization for the
inhomogeneities. The initial singularity is resolved
in this inhomogeneous model in an extremely simple way
and without imposing special boundary conditions, thus
ensuring the robustness and generality of this
resolution. Furthermore this quantization constitutes
a well founded step towards the extraction of physical
results and consequences from loop quantum cosmology,
given the central role of the inhomogeneities in
modern cosmology.
\end{abstract}


\maketitle

Loop quantum cosmology (LQC) \cite{lqc} has recently
undergone important developments. The study of the
physical consequences of loop quantum gravity (LQG)
\cite{lqg} for cosmology has motivated the analysis of
symmetry reduced models that can be fully quantized by
techniques that closely resemble those of LQG
\cite{abl}. Various homogeneous and isotropic models
have been successfully quantized \cite{iso}. Other
less symmetric systems have been also analyzed
\cite{chio,chi2,bane} with partially satisfactory
results.

Any realistic attempt to extract physical results from
LQC must consider inhomogeneities, which play a
central role in current cosmology. As a most suitable
arena to progress in this direction, we will study the
quantization of the linearly polarized Gowdy $T^3$
model \cite{gowd}. This family of spacetimes is highly
symmetric, but preserves one important feature of the
full theory: it has an infinite number of degrees of
freedom. Its classical solutions are indeed well known
and generically present an initial singularity
\cite{mon}. Besides, its quantization has been
addressed since long time ago by means of standard
non-polymeric techniques \cite{qGow} (see also
\cite{Kuc} for a related system, namely the
cylindrical waves). More recently, a rigorous Fock
quantization of this model has been accomplished,
which is shown to be essentially unique \cite{men}.
Nonetheless, even this Fock quantization (performed
after deparametrizing the system) fails to resolve the
cosmological singularity.

To overcome this problem, we will carry out a hybrid
quantization, that combines the polymeric quantization
characteristic of LQC applied to the homogeneous
solutions (which describe Bianchi I spacetimes) with
Fock quantization of the inhomogeneities. This
approach investigates the effects on quantum geometry
underlying LQC only on the homogeneous sector, while
disregards the discreteness of the geometry encoded by
the inhomogeneities. A most natural treatment for the
inhomogeneities is then the Fock quantization. Indeed,
one would expect that a quantum field theory for the
inhomogeneities, which can be regarded as a field
living on a homogeneous (Bianchi I) background, be
approximately valid on the polymerically quantized
background. An additional motivation for this approach
is the possibility that the inhomogeneities can be
finally considered as perturbations. We will represent
the quantum constraints of this model, find their
solutions, construct the physical Hilbert space, and
recover the conventional quantum field theory. To our
knowledge this is the first time that this task has
been carried out to this level of completion in a
system with an infinite number of physical degrees of
freedom.

With this aim, we will reduce the classical phase
space imposing gauge fixing conditions but, unlike in
\cite{men}, we will not deparametrize the system
completely, so that we will be left with a remanent
Hamiltonian constraint, namely the spatial average of
the full Hamiltonian constraint. Its quantum
counterpart is not entirely derived from LQG, since
the inhomogeneities are not quantized \`{a} la loop,
but it retains interesting polymeric features. In
particular, we will face the problem of polymerically
quantizing the internal time. As we will see this may
affect the conventional concept of evolution.

We will also analyze the role played by the
inhomogeneities in the quantum theory and thus the
robustness of the results obtained in homogeneous LQC.
In fact, as we have commented, a fundamental
motivation for this hybrid quantization is the
resolution of the cosmological singularity. The first
studies in homogeneous LQC already indicate that the
section of constant internal time corresponding to the
initial singularity can be avoided in the quantum
theory \cite{boj,abl}. In this letter we will show
that the loop quantization of the homogeneous degrees
of freedom, which correspond to global variables,
suffices to ``cure'' the singularity, without the need
to resort to the BKL (Belinsky, Khalatnikov, and
Lifshitz) conjecture \cite{bkl}. Therefore we do not
have to consider all the points individually. We will
see that the initial singularity disappears from our
quantum theory. Actually, nontrivial physical states
contain no contribution corresponding to the singular
initial section. Furthermore, they do not even evolve
to another branch of the universe with opposite
(triad) orientation, and can thus be viewed as arising
from ``nothing'', without imposing any particular
boundary condition.

The considered Gowdy model is a globally hyperbolic
vacuum spacetime, with spatial sections homeomorphic
to a 3-torus and two hypersurface-orthogonal spatial
Killing fields \cite{gowd}. It can be described in
terms of coordinates $\{t,\theta,\sigma,\delta\}$
adapted to the symmetries, where $\partial_\sigma$ and
$\partial_\delta$ are the two Killing fields, with
$\sigma,\delta\in S^1$. The induced spatial 3-metric,
the densitized lapse function
${N_{_{_{\!\!\!\!\!\!\sim}}\;}}$, and the shift vector
$N^{i}$, with $i\in\{\theta, \sigma,\delta\}$, depend
on $t$ and $\theta\in S^1$, and are then expandable in
Fourier series.

Suitable gauge fixing conditions for the
diffeomorphisms generated by the two Killing fields,
together with the linear polarization condition, lead
to a diagonal 3-metric. It can be characterized by
three fields that describe the norm of one of the
Killing vectors, the area of the isometry group
orbits, and the scale factor of the metric induced on
the set of group orbits. We further impose that the
generator of the conformal transformations of this
latter metric and the area of the isometry group
orbits be homogeneous functions. These conditions fix
the gauge freedom associated to the nonzero Fourier
modes of the $\theta$-momentum constraint and of the
densitized Hamiltonian constraint, and imply that the
functions $N^\theta$ and
${N_{_{_{\!\!\!\!\!\!\sim}}\;}}$ are homogeneous
\cite{men}. The classically reduced system is then
described by three canonical pairs of variables, which
correspond to the homogeneous degrees of freedom of a
Bianchi I spacetime with $T^3$-topology, and by the
infinite number of nonzero (inhomogeneous) modes of
the field unaffected by the gauge fixing, together
with their corresponding conjugate momenta (we adopt
the same parametrization as in \cite{men}). We will
call these the homogeneous and inhomogeneous sectors
respectively. There remain two constraints: the zero
mode of the $\theta$-momentum constraint and the zero
mode of the densitized Hamiltonian constraint, which
generate $S^1$ and time translations.

The homogeneous sector of this reduced Gowdy model is
described in Ashtekar variables. We adopt a diagonal
gauge for these variables, corresponding to Bianchi I
degrees of freedom. Since they can be defined in such
a way that they do not depend on the selected fiducial
metric \cite{chi2}, we choose the Euclidean one for
simplicity. The natural coordinate cell of the system
is the $T^3$-cell. Then the nontrivial components of
the $SU(2)$ gravitational connection and of the
densitized triad are ${c^{i}}/{(2\pi)}$ and
${p_{i}}/{(4\pi^2)}$ respectively, with
$\{c^i,p_j\}=8\pi G\gamma\delta^i_j$. Here, $\gamma$
is the Immirzi parameter.  We note that $p_\theta$ is
proportional to the time variable chosen in \cite{men}
to deparametrize the system.

The kinematical Hilbert space for the homogeneous
sector is constructed adapting LQG procedures to
symmetry reduced systems \cite{abl,chio}. The
elementary homogeneous variables are holonomies along
edges of oriented coordinate length $2\pi\mu_i$ in the
$i$ direction, $\mu_i$ being a real number, and triad
fluxes through rectangles orthogonal to those
directions. The configuration algebra, denoted by
$\text{Cyl}_S$, is the algebra of almost periodic
functions generated by the exponentials $\mathcal
N_{\mu_j}(c^j)=\exp(i\mu_{j}c^{j}/2)$ \cite{F1}, which
in the Dirac ket notation become $|\mu_j\rangle$. The
homogeneous sector of the kinematical Hilbert space is
the Cauchy completion of $\text{Cyl}_S$ with respect
to the discrete inner product
$\langle\mu_i|\mu_i^\prime\rangle=\delta_{\mu_i
\mu_i^\prime}$ for each direction. The basis states
$|\mu_i\rangle$ are eigenstates of the momentum
operator $\hat p_i$, while the action of
$\hat{\mathcal N}_{\mu_i^\prime}$ shifts their label
$\mu_i$.

Remember that in LQG physical areas are discrete and
have a minimum nonzero eigenvalue $\Delta$. This
feature has an imprint in LQC. The coordinate length
of the holonomy along each edge exhibits a
state-dependent minimum value $2\pi\bar\mu_i$
\cite{PVA}. Here, we will adopt the (operator)
relation ${\bar\mu_i}^2 \widehat{|p_i|}= \Delta$, as
proposed in \cite{chio} (for additional details on
this relation, see \cite{mmp}). As a result, the
operator $\hat{\mathcal N}_{\bar\mu_i}$ generates a
state-dependent minimum shift. It is convenient to
rename the states by reparametrizing the labels
$\mu_i$ so that this minimum shift becomes uniform.
Carrying out this affine reparametrization
\cite{iso,chio} one introduces, for each direction, a
new label $v_i(\mu_i)$ such that $\hat{\mathcal
N}_{\bar\mu_i}|v_i\rangle=|v_i+1\rangle$ and $\hat
p_i|v_i\rangle=3^{1/3}\Delta\,
\text{sgn}(v_i)|v_i|^{2/3}|v_i\rangle$. The operators
acting on the homogeneous sector can then be written
in terms of these elementary ones, which are densely
defined in the domain spanned by the states
$|v_i\rangle$.

For the inhomogeneous modes of the only field present
in the theory, we adopt the Fock quantization of
\cite{men}. We represent this field in terms of the
basis of $n$-particle states naturally associated to a
free massless scalar field \cite{men},
$|\{n_m\}\rangle:=|...,n_{-m},...,n_m,...\rangle$,
where $m\neq0$ and $n_m<\infty$ is the occupation
number of the $m$-th mode. In these states, only a
finite set of these occupation numbers differ from
zero. The proper Fock subspace annihilated by the
constraint that generates $S^1$-translations will be
called $\mathcal F_p$. A basis for it is provided by
the $n$-particle states that verify the condition
$\sum_{m>0}m(n_m-n_{-m})=0$.

In order to quantize the densitized Hamiltonian
constraint of the model, let us first focus on Bianchi
I.

LQG techniques provide the (nondensitized) quantum
Hamiltonian constraint $\widehat{C}_{\text{BI}}$ for
Bianchi I in terms of the elementary operators $\hat
p_i$ and $\hat{\mathcal N}_{\mu_i}$
\cite{chio,mmp,merce}. We adopt a factor ordering such
that this constraint is a sum of terms factorized in
each direction, each of these factors being symmetric.
Besides, our prescriptions ensure that
$\widehat{C}_{\text{BI}}$ annihilates the proper
subspace on which any of the operators $\hat{p_i}$
vanishes \cite{mmp,merce}. Taking into account that
the volume operator is $\hat V=\sqrt{|\hat
p_\theta\hat p_\sigma\hat p_\delta|}$, we will call it
the subspace of zero-volume states. Remarkably, this
subspace decouples when the constraint is imposed
because, in addition, its complement is invariant
under the action of $\widehat{C}_{\text{BI}}$.
Therefore, we will restrict to this complement in the
following.

We are now ready to write down the densitized quantum
Hamiltonian constraint
$\widehat{\mathcal{C}}_{\text{BI}}$ for Bianchi I. LQG
procedures give the regularized inverse volume
operator $\widehat{1/V}$. It only annihilates the
zero-volume states, which are removed from our theory.
Consequently, its inverse $[\widehat{1/V}]^{-1}$ is
well defined. In principle, Bianchi I physical states
(i.e. those annihilated by $\widehat{C}_{\text{BI}}$)
will not be normalizable in the kinematical Hilbert
space. These states, denoted by $(\psi|$, live in a
larger space \cite{F2} on which we can define
\[
\widehat{\mathcal{C}}_{\text{BI}}=
[\widehat{1/V}]^{-1/2} \widehat{C}_{\text{BI}}
[\widehat{1/V}]^{-1/2}.
\]
This operator annihilates the transformed physical
states $({\tilde\psi}|$, related to $(\psi|$ by the
bijection $(\tilde{\psi}|=(\psi|
[\widehat{1/V}]^{1/2}$.

In the full Gowdy model, once the homogeneity of the
densitized lapse function is taken into account, the
densitization of the Hamiltonian constraint is carried
out by means of the transformation explained above.
Again, states on the kernel of any of the $\hat p_i$'s
are decoupled. The densitized Hamiltonian constraint
$\widehat{\mathcal{C}}_{\text{G}}$ is the sum of two
terms: the already quantized Bianchi I part and the
part that involves the inhomogeneities. The latter
depends on the absolute value of our internal time
variable, $|p_\theta|$. In particular, there is a term
quadratic in the inverse of its square root, for which
LQC provides the corresponding regularized operator
(which acts diagonally on the states
$|v_\theta\rangle$). In addition, this part of the
constraint also depends on the connections, which are
not well defined operators in LQC. This difficulty can
be overcome if we notice that the only dependence on
the connections occurs through factors of the form
$c^ip_i$. Indeed, if we compare
$\widehat{\mathcal{C}}_{\text{BI}}$ with its classical
counterpart, it is possible to identify a natural
quantum prescription to represent $c^ip_i$. The
corresponding symmetric operator, $\widehat\Theta_i$,
can be written in terms of the operators $|\hat p_i|$,
$\text{sgn}(\hat p_i)$, and $\hat{\mathcal
N}_{\pm2\bar\mu_i}$. It can be shown to be essentially
self-adjoint. Its action on the states $|v_i\rangle$
is of the form \cite{mmp,merce}
\[\widehat{\Theta}_i|v_i\rangle=-i\pi\gamma
l_{\text p}^2\left[f_+(v_i)|v_i
+2\rangle-f_-(v_i)|v_i-2\rangle\right],\] where
$l_{\text p}=\sqrt{G\hbar}$ is the Planck length and
the functions $f_+(v_i)$ and $f_-(v_i)$ possess the
relevant property that they vanish in the intervals
$[-2,0]$ and $[0,2]$, respectively.

The explicit expression of the constraint
$\widehat{\mathcal{C}}_{\text{G}}$ is \cite{merce}
\begin{align*}
\widehat{\mathcal{C}}_{\text{G}}
 &=-\frac{2}{\gamma^2}
 \bigg[\widehat{\Theta}_\theta
\widehat{\Theta}_\delta+
\widehat{\Theta}_\theta\widehat{\Theta}_\sigma+
\widehat{\Theta}_\sigma
\widehat{\Theta}_\delta\bigg]\\
 &+l_{\text p}^2\bigg[\frac{(\widehat{\Theta}_\sigma
 +\widehat{\Theta}_\delta)^2}{\gamma^2}
\bigg(\widehat{\frac{1}{\sqrt{|p_\theta|}}}\bigg)^2
\widehat{H}_\text{int}^\xi+32\pi^2\widehat{|p_\theta|}
\widehat{H}_0^\xi\bigg],\\
\widehat{H}_\text{int}^\xi
 &=\sum_{m\neq
0}\frac{1}{2|m|}\left[2\hat{a}^{\dagger}_m\hat{a}_m+
\hat{a}^{\dagger}_m\hat{a}^{\dagger}_{-m}+\hat{a}_m
\hat{a}_{-m}\right].
\end{align*}
Here,
$\widehat{H}_0^\xi=\sum_{m}|m|\hat{a}^{\dagger}_m
\hat{a}_m$ is the Hamiltonian of a massless free
scalar field, and $\hat{a}_m$ and $\hat{a}_m^\dagger$
are the annihilation and creation operators of
particles in the $m$-th mode corresponding to that
field, such that $[\hat{a}_m,\hat{a}_{\tilde
m}^\dagger]=\delta_{m\tilde m}$. In the above formula
for $\widehat{\mathcal{C}}_{\text{G}}$, the first line
is the densitized Hamiltonian constraint
$\widehat{\mathcal{C}}_{\text{BI}}$ of Bianchi I.

Since $\widehat\Theta_i$ is a difference operator and
actually it does not relate states $|v_i\rangle$ with
$v_i>0$ to states with $v_i<0$, the label $v_i$ can be
restricted to any of the semilattices $\mathcal
L_{\varepsilon_i}^\pm=\{\pm(\varepsilon_i+2k),
k\in\mathbb{N}\}$, where $\varepsilon_i\in(0,2]$.
Semilattices corresponding to different
$\varepsilon_i$ or different signs do not get mixed
under the action of the constraint. In this sense the
homogeneous sector is superselected. We can then focus
our study on the kinematical Hilbert space
$(\otimes_i\mathcal H_{\varepsilon_i})\otimes\mathcal
F_p$, where $\mathcal H_{\varepsilon_i}$ is the
completion of the span of states $|v_i\rangle$ with
$v_i$ in the specific semilattice $\mathcal
L_{\varepsilon_i}^+$.

Note also that  $\widehat\Theta_\sigma$ and
$\widehat\Theta_\delta$ are Dirac observables. A
careful analysis, complemented with numerical
evidence, supports that the spectrum of
$\widehat\Theta_i$ is absolutely continuous and
coincides with the real line \cite{mmp}. Therefore its
generalized eigenstates $|\lambda_i\rangle$ are
delta-normalizable with respect to the Lebesgue
measure $d\lambda_i$.

The constraint $\widehat{\mathcal{C}}_{\text{G}}$ is a
well defined operator in the domain spanned by
$\{|v_\theta\rangle\otimes|v_\sigma\rangle\otimes
|v_\delta\rangle\otimes|\{n_m\}\rangle; v_i\neq0\}$
\cite{new}. This result shows the viability of
combining the polymeric and Fock quantizations.
Furthermore, an analysis of the associated deficiency
index equation indicates that it possesses no
normalizable solutions and therefore that the
constraint operator is essentially self-adjoint. Thus,
group averaging techniques can in principle be
available to determine the space of solutions to the
Hamiltonian constraint.

Nonetheless, one can formally obtain the generic
expression of the solutions without the need to resort
to group averaging, whose application is not immediate
in this case. For convenience, we will use the
(generalized) eigenstates of $\hat p_\theta$,
$\widehat\Theta_\sigma$, and $\widehat\Theta_\delta$,
instead of those of the $\hat p_i$'s, to decompose the
homogeneous sector. Then, if we formally expand the
states $(\tilde\psi|$, we can represent them by the
coefficients $(\tilde\psi_{\lambda_\sigma,
\lambda_\delta}(v_\theta)|\{n_m\}\rangle$.
$\widehat\Theta_\theta$ is the only operator in
$\widehat{\mathcal{C}}_{\text{G}}$ that does not act
diagonally on the homogeneous sector. Consequently,
the Hamiltonian constraint becomes a recurrence
equation that relates the coefficients in the three
consecutive sections $(v_\theta-2)$, $v_\theta$, and
$(v_\theta+2)$. However the coefficient evaluated at
$(v_\theta-2)$ appears multiplied by the factor
$f_-(v_\theta)$. Since $f_-(\varepsilon_\theta)=0$,
the recurrence equation turns into a consistency
relation for the coefficients in the two first
sections $\varepsilon_\theta$ and
$(\varepsilon_\theta+2)$. As a result any coefficient
$(\tilde\psi_{\lambda_\sigma,\lambda_\delta}(v_\theta)
|\{n_m\}\rangle$ can be expressed in terms of the
initial ones
$(\tilde\psi_{\lambda_\sigma,\lambda_\delta}
(\varepsilon_\theta)|\{n_{\tilde{m}}\}\rangle$. The
formal solution of the constraint is, obviating the
$\lambda$-dependence and the projection on
$n$-particle states, \begin{align*}
\big(\tilde\psi(\varepsilon_\theta+2k)\big|
&=\big(\tilde\psi(\varepsilon_\theta)\big|
\sum_{O(k)}\Big[
\prod_{\{r_i\}}F(\varepsilon_\theta+2r_i+2)\Big]\\
&\times\mathcal P \Big[\prod_{\{s_j\}}
\widehat{H}^\xi(\varepsilon_\theta+2s_j)\Big],
\end{align*}
where $O(k)$ is the set of possible ways to move from
0 to $k$ by jumps of one or two steps and, for each
element of $O(k)$, $\{s_j\}$ is the set of integers
followed by a jump of one step, while $\{r_i\}$ are
those followed by a jump of two steps \cite{new}.
Besides, $F(v_{\theta})=f_-(v_{\theta})/
f_+(v_{\theta})$, the symbol $\mathcal P$ denotes path
ordering, and $\widehat{H}^\xi(v_{\theta})$ has the
form:
\[\widehat{H}^\xi(v_{\theta})=y_1(v_{\theta})+
y_2(v_{\theta})\widehat{
H}_\text{int}^\xi+y_3(v_{\theta})\widehat{H}_0^\xi,\]
$y_k(v_{\theta})$ being $\lambda$-dependent functions
whose detailed expression is irrelevant for our
discussion.

As we have seen,
$(\tilde\psi_{\lambda_\sigma,\lambda_\delta}
(\varepsilon_\theta)|$ completely determines the
solution, so that we can identify the latter with the
corresponding data on the initial section
$v_\theta=\varepsilon_\theta$. To endow the space of
these solutions with a Hilbert structure, we choose a
complete set of real classical observables and require
that their quantum counterparts be self-adjoint
operators. At this stage it is easier to relax the
$S^1$-gauge symmetry, so that we can choose the
operators that represent the Fourier sine and cosine
coefficients of the nonzero modes as the basic
observables for the inhomogeneous sector. For the
homogeneous sector, we can select as a complete set of
observables (in the $\lambda$-representation) the
operators of multiplication by $\lambda_{\sigma}$ and
$\lambda_{\delta}$ (i.e., $\widehat\Theta_\sigma$ and
$\widehat\Theta_\delta$), together with the respective
derivatives evaluated on the considered section, $-i
\partial_{\lambda_{\sigma}}|_{\varepsilon_\theta}$ and
$-i\partial_{\lambda_{\delta}}|_{\varepsilon_\theta}$.
In order to obtain the physical Hilbert space we first
demand reality conditions on all these observables and
then restore the $S^1$-symmetry. The final result is
that, for generic initial data
$(\tilde\psi_{\lambda_\sigma,\lambda_\delta}
(\varepsilon_\theta)|$, the physical Hilbert space
that one obtains in this way is unitarily equivalent
to $L^2(\mathbb{R}^2,d\lambda_\sigma
d\lambda_\delta)\otimes\mathcal F_p$. In particular,
let us emphasize that we then recover the Fock
quantization of \cite{men} for the physical
inhomogeneous sector.

Several comments are in order once we have completed
the quantization of this Gowdy model.

We remark that the general expression of the solutions
is only formal. A more rigorous interpretation in
terms of structures associated to the kinematical
Hilbert space is obstructed by the production of
infinite particle pairs caused by the operator
$\widehat{H}_\text{int}^\xi$ and by the
$v_{\theta}$-dependence of the inhomogeneous part of
the constraint, which prevents its diagonalization in
terms of particle states with a fixed mass for all the
$v_{\theta}$-sections. Nevertheless we can in fact
make full sense of the solutions in certain
situations.

This is the case if one introduces a cut-off for the
wavenumber of the field modes, what can be understood
as a further symmetry reduction of the Gowdy model.
Solutions to the constraint are then in the algebraic
dual of the dense set of $n$-particle states on each
$v_{\theta}$-section. Actually, the mentioned physical
Hilbert space for the full Gowdy model can then be
attained in the limit in which the cut-off is removed.
A related possibility that we will explore in a future
research is to introduce a dynamical cut-off,
dependent on the value of $v_\theta$, as an effective
way to account for the minimum coordinate length
$\bar{\mu}_{\theta}$.

Other situations in which the structure of the
solutions is well under control is when one regards
the inhomogeneous part of the Hamiltonian constraint
as a perturbation around Bianchi I and truncates the
solutions at second perturbative order in it or,
alternatively, if one views as a perturbation only the
term that contains $\widehat{H}_\text{int}^\xi$ and
makes the truncation at first order. Then the
solutions can be defined again, up to the analyzed
perturbative order, in the dual of the $n$-particle
states for each $v_{\theta}$-section and the physical
Hilbert space that one obtains is just
$L^2(\mathbb{R}^2,d\lambda_\sigma
d\lambda_\delta)\otimes\mathcal F_p$.

Note that in the above cases we can adopt the picture
that solutions are initial data on the section
$v_\theta=\!\varepsilon_\theta$ which evolve to new
data on any other section
$v_\theta=\varepsilon_\theta+2k$. Identifying
sections, we arrive in principle at a concept of
evolution for physical states which would not be
unitary (i.e. the norm would not be preserved) as a
consequence of the polymeric quantization of the
internal time.

Furthermore, the loop quantization carried out on the
homogeneous sector allows us to get insight about the
resolution of the cosmological singularity. The
classical Gowdy model generically presents an initial
singularity where some of the variables $p_i$ vanish.
Since zero-volume states are removed in the quantum
theory, the kernel of any of the operators $\hat p_i$
is empty. Therefore the singularity disappears already
at the kinematical level, and, at least in this sense,
is resolved. Note that, in a standard quantization of
the homogeneous sector, the operators $\hat p_i$ have
instead a continuous spectrum. Thus, their kernel is
not a proper subspace which could be decoupled in
order to remove the singularity in the kinematical
Hilbert space. Moreover, the singularity persists then
also in the physical Hilbert space. This fact can be
easily realized in the completely deparametrized
system quantized \`{a} la Fock \cite{men}, since the
domain of definition of the time parameter plays there
the role of the spectrum of $|\hat{p}_{\theta}|$
quantized in the standard way. Curvature invariants
become explicitly time-dependent quantum observables
that indeed diverge in the limit when the time
parameter vanishes \cite{mon}.

To conclude, we have successfully quantized this
inhomogeneous Gowdy model by combining loop and Fock
quantizations. The attained results prove that our
hybrid quantization is feasible and well posed. More
precisely, we have represented rigorously the quantum
constraints and explicitly found their solutions,
which have been endowed with a Hilbert structure. The
physical Hilbert space thus obtained reproduces the
conventional Fock quantization of the inhomogeneous
modes. As far as we know, this is the first
quantization carried out to this degree of completion
in an inhomogeneous cosmological system using
polymeric techniques. The loop quantization of the
homogeneous sector proves enough to resolve the
classical singularity, inasmuch as it disappears from
the quantum theory. In fact, it is not only avoided in
physical states, but they never cross it. The quantum
states of our Gowdy universe start in a single
nonsingular section, without requiring any boundary
conditions. This result reinforces the robustness of
the big-bang avoidance in LQC and paves the road for
the extension of the singularity resolution to
inhomogeneous cases.

\acknowledgements The authors are very grateful to
J.M. Velhinho and T. Pawlowski. This work was
supported by the Spanish Grants FIS2005-05736-C03-02
(its continuation FIS2008-06078-C03-03),
FIS2006-26387-E, and CSD2007-00042 (CPAN); and M.M-B.
by CSIC and the European Social Fund under the grant
I3P-BPD2006.


\begin{thebibliography}{99}

\bibitem{lqc} M. Bojowald, Living Rev. Rel. {\bf11}, 4
(2008).

\bibitem{lqg} T. Thiemann, {\it{Modern Canonical Quantum
General Relativity}} (CUP, Cambridge, UK, 2007).

\bibitem{abl} A. Ashtekar, M. Bojowald, and J. Lewandowski,
Adv. Theor. Math. Phys. {\bf7}, 233 (2003).

\bibitem{iso} See, e.g., A. Ashtekar, T. Pawlowski, and P.
Singh, Phys. Rev. Lett. {\bf96}, 141301 (2006); Phys.
Rev. D {\bf73}, 124038 (2006); {\bf74}, 84003 (2006).

\bibitem{chio} D.W. Chiou, Phys. Rev. D {\bf75}, 24029 (2007).

\bibitem{chi2} D.W. Chiou, Phys. Rev. D {\bf76}, 124037 (2007).

\bibitem{bane} K. Banerjee and G. Date, Class. Quantum Grav.
{\bf25}, 145004 (2008).

\bibitem{gowd} R.H. Gowdy, Ann. Phys. {\bf83}, 203 (1974).

\bibitem{mon} V. Moncrief, Phys. Rev. D {\bf23}, 312 (1981).

\bibitem{qGow} See, e.g., C.W. Misner, Phys. Rev. D {\bf8}, 3271
(1973); B.K. Berger, Ann. Phys. {\bf83}, 458 (1974); Phys. Rev. D
{\bf 11}, 2770 (1975); Ann. Phys. {\bf 156}, 155 (1984); G.A. Mena
Marug\'{a}n, Phys. Rev. D {\bf56}, 908 (1997); M. Pierri, Int. J.
Mod. Phys. Phys. D {\bf11}, 135 (2002).

\bibitem{Kuc} See, e.g., K. Kuchar, Phys. Rev. D {\bf4}, 955 (1971);
A. Ashtekar and M. Pierri, J. Math. Phys. {\bf37}, 6250 (1996); J.F.
Barbero G., G.A. Marug\'{a}n, and E.J.S. Villase\~nor, Phys. Rev. D
{\bf67}, 124006 (2003).

\bibitem{men} A. Corichi, J. Cortez, and G.A. Mena
Marug\'{a}n, Phys. Rev. D {\bf73}, 41502 (2006);
{\bf73}, 84020 (2006); A. Corichi, J. Cortez, G.A.
Mena Marug\'{a}n, and J.M. Velhinho, Class. Quantum
Grav. {\bf23}, 6301 (2006); J. Cortez, G.A. Mena
Marug\'{a}n, and J.M. Velhinho, Phys. Rev. D {\bf75},
84027 (2007).

\bibitem{boj} M. Bojowald,  Phys. Rev. Lett. {\bf86}, 5227
(2001).

\bibitem{bkl} V.A. Belinsky, I.M. Khalatnikov, and E.M.
Lifshitz, Adv. Phys. {\bf31}, 639 (1982).

\bibitem{F1} We will not use the Einstein summation
convention.

\bibitem{PVA} There exist different proposals to define
this minimum value, which are still under discussion.
This issue has been considered in \cite{chi2} and by
A. Ashtekar (unpublished).

\bibitem{mmp} M. Mart\'{\i}n-Benito, G.A. Mena
Marug\'{a}n, and T. Paw\-lowski, arXiv:0804.3157
[Phys. Rev. D (to be published)].

\bibitem{merce} M. Mart\'{\i}n-Benito, {\it
{Cuantizaci\'on del Modelo de Gowdy $T^3$ con
Polarizaci\'on Lineal}} (Master thesis, Univ.
Complutense de Madrid, 2007).

\bibitem{F2} Namely, the algebraic dual of the subset
of $\text{Cyl}_S$ of nonzero-volume states.

\bibitem{new} L.J. Garay, M. Mart\'{\i}n-Benito, G.A.
Mena Marug\'an, and J.M. Velhinho (unpublished).

\end{thebibliography}
\end{document}